# Generating GHZ state in $2m$-qubit spin network

M. A. Jafarizadeh$^{a,d}$ *, R. Sufiani$^{a}$ †, F. Eghbalifam$^{a}$, M. Azimi$^{a}$,

S. F. Taghavi$^{c}$ and E. Barati$^{e}$,

$^{a}$Department of Theoretical Physics and Astrophysics, University of Tabriz, Tabriz 51664, Iran.

$^{b}$Center of excellence for photonic, University of Tabriz, Tabriz 51664, Iran.

$^{c}$Institute for Studies in Theoretical Physics and Mathematics, Tehran 19395-1795, Iran.

$^{d}$Research Institute for Fundamental Sciences, Tabriz 51664, Iran.

$^{e}$Institute of Physical Chemistry, Polish Academy of Sciences, Kasprzaka 44/52, 01-224 Warszawa, Poland.

August 11, 2018

*E-mail:jafarizadeh@tabrizu.ac.ir

†E-mail:sofiani@tabrizu.ac.ir

**Abstract**

We consider a pure $2m$-qubit initial state to evolve under a particular quantum mechanical spin Hamiltonian, which can be written in terms of the adjacency matrix of the Johnson network $J(2m,m)$. Then, by using some techniques such as spectral distribution and stratification associated with the graphs, employed in [1, 2], a maximally entangled $GHZ$ state is generated between the antipodes of the network. In fact, an explicit formula is given for the suitable coupling strengths of the hamiltonian, so that a maximally entangled state can be generated between antipodes of the network. By using some known multipartite entanglement measures, the amount of the entanglement of the final evolved state is calculated, and finally two examples of four qubit and six qubit states are considered in details.

**Keywords: maximal entanglement , GHZ states, Johnson network, Stratification, Spectral distribution**

**PACs Index: 01.55.+b, 02.10.Yn**

# 1 Introduction

The idea to use quantum spin chains for short distance quantum communication was put forward by Bose [3]. After the work of Bose, the use of spin chains [4]-[18] and harmonic chains [19] as quantum wires have been proposed. In the previous work [1], the so called distance-regular graphs have been considered as spin networks (in the sense that with each vertex of a distance-regular graph, a qubit or a spin was associated) and perfect state transfer (PST) of a single qubit state over antipodes of these networks has been investigated. In that work, a procedure for finding suitable coupling constants in some particular spin Hamiltonians has been given so that perfect and optimal transfer of a quantum state between antipodes of the corresponding networks can be achieved, respectively.

Entanglement is one of the other important tasks in quantum communication. Quantum entanglement in spin systems is an extensively-studied field in recent years [20,21,22,23], in the advent of growing realization that entanglement can be a resource for quantum information processing. Within this general field, entanglement of spin1/2 degrees of freedom, qubits, has been in focus for an obvious reason of their paramount importance for quantum computers, not to mention their well-known applicability in various condensed-matter systems, optics and other branches of physics. In [24], authors attempted to generate a Bell state between distant vertices in a permanently coupled spin network interacting via invariant stratification graphs (ISGs). At the first step, they established an upper bound over achievable entanglement between the reference site and the other vertices. Due to this upper bound they found that creation of a Bell state between the reference site and a vertex is possible if the stratum of that vertex is a single element, e.g. antipodal ISGs. The present work focuses on the provision of $GHZ$ state, by using a $2m$-qubit initial product state. To this end, we will consider the Johnson networks $J(2m, m)$ (which are distance-regular) as spin networks. Then, we use the algebraic properties of these networks in order to find suitable coupling constants in some particular spin Hamiltonians so that $2m$-qubit $GHZ$ state can be achieved.

The organization of the paper is as follows: In section 2, we review some preliminary facts about graphs and their adjacency matrices, spectral distribution associated with them; In particular, some properties of the networks derived from symmetric group $S_n$ called also Johnson networks are reviewed. Section 3 is devoted to $2m$-qubit $GHZ$ state provision by using algebraic properties of Johnson network $J(2m, m)$, where a method for finding suitable coupling constants in particular spin Hamiltonians is given so that maximal entanglement in final state is possible. The paper is ended with a brief conclusion and two appendices.

## 2 preliminaries

### 2.1 Graphs and their adjacency matrices

A graph is a pair $\Gamma = (V, E)$, where $V$ is a non-empty set called the vertex set and $E$ is a subset of $\{(x, y) : x, y \in V, x \neq y\}$ called the edge set of the graph. Two vertices $x, y \in V$ are called adjacent if $(x, y) \in E$, and in that case we write $x \sim y$. For a graph $\Gamma = (V, E)$, the adjacency matrix $A$ is defined as

$$(A)_{\alpha,\beta} = \begin{cases} 1 & \text{if } \alpha \sim \beta \\ 0 & \text{otherwise} \end{cases}. \tag{2-1}$$

Conversely, for a non-empty set $V$, a graph structure is uniquely determined by such a matrix indexed by $V$. The degree or valency of a vertex $x \in V$ is defined by

$$\kappa(x) = |\{y \in V : y \sim x\}| \tag{2-2}$$

where, $|\cdot|$ denotes the cardinality. The graph is called regular if the degree of all of the vertices be the same. In this paper, we will assume that graphs under discussion are regular. A finite sequence $x_0, x_1, ..., x_n \in V$ is called a walk of length $n$ (or of $n$ steps) if $x_{i-1} \sim x_i$ for all $i = 1, 2, ..., n$. Let $l^2(V)$ denote the Hilbert space of $C$-valued square-summable functions on $V$. With each $\beta \in V$ we associate a vector $|\beta\rangle$ such that the $\beta$-th entry of it is 1 and all of the other entries of it are zero. Then $\{|\beta\rangle : \beta \in V\}$ becomes a complete orthonormal basis of $l^2(V)$. The adjacency matrix is considered as an operator acting in $l^2(V)$ in such a way that

$$A|\beta\rangle = \sum_{\alpha \sim \beta} |\alpha\rangle. \tag{2-3}$$

## 2.2 Spectral distribution associated with the graphs

Now, we recall some preliminary facts about spectral techniques used in the paper, where more details have been given in Refs. [26,27,28,29]

Actually the spectral analysis of operators is an important issue in quantum mechanics, operator theory and mathematical physics [30,31]. As an example $\mu(dx) = |\psi(x)|^2 dx$ ($\mu(dp) = |\widetilde{\psi}(p)|^2 dp$) is a spectral distribution which is assigned to the position (momentum) operator $\hat{X}(\hat{P})$. The mathematical techniques such as Hilbert space of the stratification and spectral techniques have been employed in [32,33] for investigating continuous time quantum walk on graphs. Moreover, in general quasi-distributions are the assigned spectral distributions of two hermitian non-commuting operators with a prescribed ordering. For example the Wigner distribution in phase space is the assigned spectral distribution for two non-commuting operators $\hat{X}$ (shift operator) and $\hat{P}$ (momentum operator) with Wyle-ordering among them [34, 35]. It is well known that, for any pair $(A, |\phi_0\rangle)$ of a matrix $A$ and a vector $|\phi_0\rangle$, one can assign a measure $\mu$ as follows

$$\mu(x) = \langle \phi_0 | E(x) | \phi_0 \rangle, \tag{2-4}$$

where $E(x) = \sum_i |u_i\rangle\langle u_i|$ is the operator of projection onto the eigenspace of $A$ corresponding to eigenvalue $x$, i.e.,

$$A = \int x E(x) dx. \tag{2-5}$$

Then, for any polynomial $P(A)$ we have

$$P(A) = \int P(x) E(x) dx, \tag{2-6}$$

where for discrete spectrum the above integrals are replaced by summation. Therefore, using the relations (2-4) and (2-6), the expectation value of powers of adjacency matrix $A$ over reference vector $|\phi_0\rangle$ can be written as

$$\langle\phi_0|A^m|\phi_0\rangle = \int_R x^m \mu(dx), \quad m = 0, 1, 2, .... \quad (2\text{-}7)$$

Obviously, the relation (2-7) implies an isomorphism from the Hilbert space of the stratification onto the closed linear span of the orthogonal polynomials with respect to the measure $\mu$.

## 2.3 Underlying networks derived from symmetric group $S_n$

Let $\lambda = (\lambda_1, ..., \lambda_m)$ be a partition of $n$, i.e., $\lambda_1 + ... + \lambda_m = n$. We consider the subgroup $S_m \otimes S_{n-m}$ of $S_n$ with $m \leq [\frac{n}{2}]$. Then we assume the finite set $M^\lambda = \frac{S_n}{S_m \otimes S_{n-m}}$ with $|M^\lambda| = \frac{n!}{m!(n-m)!}$ as vertex set. In fact, $M^\lambda$ is the set of $(m-1)$-faces of $(n-1)$-simplex (recall that, the graph of an $(n-1)$-simplex is the complete graph with $n$ vertices denoted by $K_n$). If we denote the vertex $i$ by $m$-tuple $(i_1, i_2, ..., i_m)$, then the adjacency matrices $A_k$, $k = 0, 1, ..., m$ are defined as

$$(A_k)_{i,j} = \begin{cases} 1 & \text{if } \partial(i,j) = k, \\ 0 & \text{otherwise} \quad (i,j \in M^\lambda) \end{cases}, \quad k = 0, 1, ..., m. \quad (2\text{-}8)$$

where, we mean by $\partial(i,j)$ the number of components that $i = (i_1, ..., i_m)$ and $j = (j_1, ..., j_m)$ are different (this is the same as Hamming distance which is defined in coding theory). The network with adjacency matrices defined by (2-8) is known also as the Johnson network $J(n,m)$ and has $m + 1$ strata such that

$$\kappa_0 = 1, \quad \kappa_l = \binom{m}{m-l}\binom{n-m}{l}, \quad l = 1, 2, ..., m. \quad (2\text{-}9)$$

One should notice that for the purpose of maximal entanglement provision, we must have $\kappa_m = 1$ which is fulfilled if $n = 2m$, so we will consider the network $J(2m, m)$ (hereafter we will take $n = 2m$ so that we have $\kappa_m = 1$). If we stratify the network $J(2m, m)$ with respect to a given reference node $|\phi_0\rangle = |i_1, i_2, ..., i_m\rangle$, where $|i_1, i_2, ..., i_m\rangle \equiv |0...0\underbrace{1}_{i_1}0...0\underbrace{1}_{i_2}0...0\underbrace{1}_{i_m}0\rangle$ and $i_1 \neq i_2 \neq ... \neq i_m$. The unit vectors $|\phi_i\rangle$, $i = 1, ..., m$ are defined as

$$|\phi_1\rangle = \frac{1}{\sqrt{\kappa_1}}(\sum_{i'_1 \neq i_1}|i'_1, i_2, ..., i_m\rangle + \sum_{i'_2 \neq i_2}|i_1, i'_2, i_3, ..., i_m\rangle + ... + \sum_{i'_m \neq i_m}|i_1, ..., i_{m-1}, i'_m\rangle),$$

$$|\phi_2\rangle = \frac{1}{\sqrt{\kappa_2}}\sum_{k \neq l=1}^{m}\sum_{i'_l \neq i_l, i'_k \neq i_k}|i_1, ...i_{l-1}, i'_l, i_{l+1}, ..., i_{k-1}, i'_k, i_{k+1}..., i_m\rangle,$$

$$\vdots$$

$$|\phi_j\rangle = \frac{1}{\sqrt{\kappa_j}}\sum_{k_1 \neq k_2 \neq ... \neq k_j=1}^{m}\sum_{i'_{k_1} \neq i_{k_1}, ..., i'_{k_j} \neq i_{k_j}}|i_1, ..., i_{k_1-1}, i'_{k_1}, i_{k_1+1}, ..., i_{k-1}, i'_{k_j}, i_{k_j+1}..., i_m\rangle,$$

$$\vdots$$

$$|\phi_m\rangle = \frac{1}{\sqrt{\kappa_m}}\sum_{i'_1 \neq i_1, ..., i'_m \neq i_m}|i'_1, i'_2, ..., i'_m\rangle. \tag{2-10}$$

Since the network $J(2m, m)$ is distance-regular, the above stratification is independent of the choice of reference node. The intersection array of the network is given by

$$b_l = (m-l)^2 \; ; \quad c_l = l^2. \tag{2-11}$$

Then, by using the Eq. (B-49), the QD parameters $\alpha_i$ and $\omega_i$ are obtained as follows

$$\alpha_l = 2l(m-l), \; l = 0, 1, ..., m; \quad \omega_l = l^2(m-l+1)^2, \; l = 1, 2, ..., m. \tag{2-12}$$

Then, one can show that [27]

$$A|\phi_l\rangle = (l+1)(m-l)|\phi_{l+1}\rangle + 2l(m-l)|\phi_l\rangle + l(m-l+1)|\phi_{l-1}\rangle. \tag{2-13}$$

## 3 $GHZ$ state generation by using quantum mechanical Hamiltonian in the network $J(2m, m)$

The model we consider is the distance-regular Johnson network $J(2m, m)$ consisting of $N = C_m^{2m} = \frac{(2m)!}{m!m!}$ sites labeled by $\{1, 2, ..., N\}$ and diameter $m$. Then, we stratify the network with respect to a chosen reference site, say 1 (the discussion about stratification has been given in appendix A; In these particular networks, the first and the last strata possess only one node, i.e., $|\phi_0\rangle = |1\rangle$ and $|\phi_m\rangle = |N\rangle$). At time $t = 0$, a $2m$-qubit state is prepared in the first (reference) site of the network. We wish to provide a maximal quantum entanglement between the state of this site and the state of the $N$-th site after a well-defined period of time, in which the corresponding network is evolved under a particular Hamiltonian.

If the network be assumed as a spin network, in which a spin-1/2 particle is attached to each vertex (node) of the network, the Hilbert space associated with the network is given by $\mathcal{H} = (C^2)^{\otimes 2m}$. The standard basis for an individual qubit is chosen to be $|0\rangle = |\downarrow\rangle, |1\rangle = |\uparrow\rangle$. Then we consider the Hamiltonian

$$H_s = \frac{1}{2} \sum_{1 \leq i < j \leq 2m} H_{ij} \tag{3-14}$$

where, $H_{ij} = \sigma_i \cdot \sigma_j$ and $\sigma_i$ is a vector with familiar Pauli matrices $\sigma_i^x, \sigma_i^y$ and $\sigma_i^z$. One can easily see that, the Hamiltoniaan (3.14) commutes with the total $z$ component of the spin, i.e., $[\sigma_{total}^z, H_s] = 0$, hence the Hilbert space $\mathcal{H}$ decompose into invariant subspaces, each of which is a distinct eigenspace of the operator $\sigma_{total}^z$. So the total number of up and down spins are invariant under action of Hamiltonian or time evolution operator. Now, we recall that the kets $|i_1, i_2, \ldots, i_{2m}\rangle$ with $i_1, \ldots, i_{2m} \in \{\uparrow, \downarrow\}$ form an orthonormal basis for Hilbert space $\mathcal{H}$. Then, one can easily obtain

$$H_{ij}|\ldots \underbrace{\uparrow}_{i} \ldots \underbrace{\uparrow}_{j} \ldots\rangle = |\ldots \underbrace{\uparrow}_{i} \ldots \underbrace{\uparrow}_{j} \ldots\rangle$$

and

$$H_{ij}|\ldots \underbrace{\uparrow}_{i} \ldots \underbrace{\downarrow}_{j} \ldots\rangle = -|\ldots \underbrace{\uparrow}_{i} \ldots \underbrace{\downarrow}_{j} \ldots\rangle + 2|\ldots \underbrace{\downarrow}_{i} \ldots \underbrace{\uparrow}_{j} \ldots\rangle. \tag{3-15}$$

Equation (3.15) implies that the action of $H_{ij}$ on the basis vectors is equivalent to the action of the operator $2P_{ij} - I$, i.e. we have

$$H_{ij} = 2P_{ij} - I \tag{3-16}$$

where $P_{ij}$ is the permutation operator acting on sites $i$ and $j$. So

$$\frac{1}{2}\sum_{1\leq i<j\leq 2m}\sigma_i\cdot\sigma_j = \sum_{1\leq i<j\leq 2m}P_{ij} - \frac{1}{2}\binom{2m}{2}I, \qquad (3\text{-}17)$$

In fact restriction of the operator $\sum_{1\leq i<j\leq 2m}P_{ij}$ on the $m$-particle subspace (subspace spanned by the states with $m$ spin up) which has dimension $C_m^{2m}$, is written as the adjacency matrix $A$ of the Johnson network $J(2m,m)$, as

$$\sum_{1\leq i<j\leq 2m}P_{ij} = A + m(m-1)I. \qquad (3\text{-}18)$$

For more details see Ref.[1]. Then we stratify the network with respect to a chosen reference site, say $|\phi_0\rangle$. At time $t=0$, the state is prepared in the $2m$-qubit state $|\psi(t=0)\rangle = |\underbrace{11\ldots 1}_{m}\rangle|\underbrace{00\ldots 0}_{m}\rangle$. Now, we consider the dynamics of the system to be governed by the Hamiltonian

$$H = \sum_{k=0}^{m} J_k P_k(1/2\sum_{1\leq i<j\leq 2m}\sigma_i\cdot\sigma_j + \frac{m}{2}I), \qquad (3\text{-}19)$$

Then, by using (3.17)-(3.20), the Hamiltonian can be written as

$$H = \sum_{k=0}^{m} J_k P_k(A) \qquad (3\text{-}20)$$

$J_k$ is the coupling strength between the reference site $|\phi_0\rangle$ and all of the sites belonging to the $k$-th stratum with respect to $|\phi_0\rangle$, and $P_k(A)$ are polynomials in terms of adjacency matrix of the Johnson network. Then, the total system is evolved under unitary evolution operator $U(t) = e^{-iHt}$ for a fixed time interval, say $t$. The final state becomes

$$|\psi(t)\rangle = \sum_{j=1}^{N} f_{jA}(t)|j\rangle \qquad (3\text{-}21)$$

where, $N$ is the number of vertices, $|j\rangle$s have $2m$ entries inclusive $m$ entries equal to 1 and the other entries are 0 and $|A\rangle = |\underbrace{11\ldots 1}_{m}\underbrace{00\ldots 0}_{m}\rangle$ so that $f_{jA}(t) := \langle j|e^{-iHt}|A\rangle$.

The evolution with the adjacency matrix $H = A \equiv A_1$ for distance-regular networks (see Appendix B) starting in $|\phi_0\rangle$, always remains in the stratification space. For distance-regular network $J(2m,m)$ for which the last stratum, i.e., $|\phi_m\rangle$ contains only one site, then

maximal entanglement between the starting site $|\phi_0\rangle \equiv |A\rangle$ and the last stratum $|\phi_m\rangle$ (the corresponding antipodal node) is generated, by choosing suitable coupling constants $J_k$. In fact, for the purpose of a maximally entangled $GHZ$ state generation between the first and the last stratum of the network, we impose the constraints that the amplitudes $\langle\phi_i|e^{-iHt}|\phi_0\rangle$ be zero for all $i = 1,...,m-1$ and $\langle\phi_0|e^{-iHt}|\phi_0\rangle = f$, $\langle\phi_m|e^{-iHt}|\phi_0\rangle = f'$. Therefore, these amplitudes must be evaluated. To do so, we use the stratification and spectral distribution associated with the network $J(2m,m)$ to write

$$\langle\phi_i|e^{-iHt}|\phi_0\rangle = \langle\phi_i|e^{-it\sum_{l=0}^{m} J_l P_l(A)}|\phi_0\rangle = \frac{1}{\sqrt{\kappa_i}}\langle\phi_0|A_i e^{-it\sum_{l=0}^{m} J_l P_l(A)}|\phi_0\rangle$$

Let the spectral distribution of the graph is $\mu(x) = \sum_{k=0}^{m} \gamma_k \delta(x-x_k)$ (see Eq. (B-53)). The Johnson network is a kind of network with a highly regular structure that has a nice algebraic description; For example, the eigenvalues of this network can be computed exactly (see for example the notes by Chris Godsil on association schemes [39] for the details of this calculation). Indeed, the eigenvalues of the adjacency matrix of the network $J(2m,m)$ (that is $x_k$'s in $\mu(x)$) are given by

$$x_k = m^2 - k(2m+1-k), \quad k = 0, 1, \ldots, m. \tag{3-22}$$

Now, from the fact that for distance-regular graphs we have $A_i = \sqrt{\kappa_i} P_i(A)$ [27], $\langle\phi_i|e^{-iHt}|\phi_0\rangle = 0$ implies that

$$\sum_{k=0}^{m} \gamma_k P_i(x_k) e^{-it\sum_{l=0}^{m} J_l P_l(x_k)} = 0, \quad i = 1,...,m-1$$

Denoting $e^{-it\sum_{l=0}^{m} J_l P_l(x_k)}$ by $\eta_k$, the above constraints are rewritten as follows

$$\sum_{k=0}^{m} P_i(x_k)\eta_k\gamma_k = 0, \quad i = 1,...,m-1,$$

$$\sum_{k=0}^{m} P_0(x_k)\eta_k\gamma_k = f$$

$$\sum_{k=0}^{m} P_m(x_k)\eta_k\gamma_k = f'. \tag{3-23}$$

From invertibility of the matrix $P_{ik} = P_i(x_k)$ (see Ref. [2]) one can rewrite the Eq. (3-23) as

$$\begin{pmatrix} \eta_0 \gamma_0 \\ \eta_1 \gamma_1 \\ \vdots \\ \eta_{d-1} \gamma_{d-1} \\ \eta_d \gamma_d \end{pmatrix} = P^{-1} \begin{pmatrix} f \\ 0 \\ \vdots \\ 0 \\ f' \end{pmatrix}. \tag{3-24}$$

The above equation implies that $\eta_k \gamma_k$ for $k = 0, 1, ..., m$ are the same as the entries in the first column of the matrix $P^{-1} = WP^t$ multiplied with $f$ and the entries in the last column multiplied with $f'$, i.e., the following equations must be satisfied

$$\eta_k \gamma_k = \gamma_k e^{-it \sum_{l=0}^{m} J_l P_l(x_k)} = (WP^t)_{k0} f + (WP^t)_{km} f' , \quad \text{for } k = 0, 1, ..., m, \tag{3-25}$$

with $W := diag(\gamma_0, \gamma_1, \ldots, \gamma_m)$. By using the fact that $\gamma_k$ and $(WP^t)_{km}$ are real for $k = 0, 1, \ldots, m$, and so we have $\gamma_k = |(WP^t)_{km}|$ and $\gamma_k = (WP^t)_{k0}$. The Eq. (3.26) can be rewritten as

$$\eta_k = e^{-it \sum_{l=0}^{m} J_l P_l(x_k)} = f + \sigma(k) f' \tag{3-26}$$

where $\sigma(k)$ is defined as

$$\sigma(k) = \begin{cases} -1 & \text{for odd } k \\ 1 & \text{otherwise} \end{cases}. \tag{3-27}$$

Assuming $f = |f|e^{i\theta}$ and $f' = |f'|e^{i\theta'}$, it should be considered $\theta' = \theta \pm \frac{\pi}{2}$ then

$$e^{-it \sum_{l=0}^{m} J_l P_l(x_k)} = e^{i\theta}(|f| \pm i\sigma(k)|f'|) = e^{i(\theta \pm \arctan(\frac{\sigma(k)|f'|}{|f|}) + 2c_k \pi)}; \quad c_k \in \mathcal{Z} \tag{3-28}$$

One should notice that, the Eq. (3.29) can be rewritten as

$$(J_0, J_1, \ldots, J_m) = -\frac{1}{t}[\theta + 2c_0\pi \pm \arctan(\frac{\sigma(k)|f'|}{|f|}), \theta + 2c_1\pi \pm \arctan(\frac{\sigma(k)|f'|}{|f|}),$$

$$, \ldots, \theta + 2c_m\pi \pm \arctan(\frac{\sigma(k)|f'|}{|f|})](WP^t) \tag{3-29}$$

or

$$J_k = -\frac{1}{t} \sum_{j=0}^{m} [\theta + 2c_j\pi \pm \arctan(\frac{\sigma(k)|f'|}{|f|})](WP^t)_{jk} \tag{3-30}$$

where $c_j$ for $j = 0, 1, \ldots, m$ are integers. The result (3-30) gives an explicit formula for suitable coupling constants so that GHZ state in the final state can be achieved. The final state is as the form

$$|\psi(t)\rangle = f|11\ldots100\ldots0\rangle + f'|00\ldots011\ldots1\rangle \tag{3-31}$$

One attempt to provide a computationally feasible and scalable quantification of entanglement in multipartite systems was made in Refs. [40,41,42]. For a pure $n$-qubit state $|\psi\rangle$, the so-called global entanglement is defined as

$$Q(|\psi\rangle) = 2(1 - \frac{1}{N}\sum_{i=0}^{N-1} Tr[\rho_i^2]) \tag{3-32}$$

where $\rho_i$ represents the density matrix of $i$th qubit after tracing out all other qubits. As seen from this definition, the global entanglement can be interpreted as the average over the (bipartite) entanglements of each qubit with the rest of the system. The global entanglement for state in Eq. (3.32) will be

$$Q(|\psi\rangle) = 4|f|^2|f'|^2 \tag{3-33}$$

Also we introduce a simple multiqubit entanglement quantifier based on the idea of bipartition and the measure negativity (which is two times the absolute value of the sum of the negative eigenvalues of the corresponding partially transposed matrix of a state $\rho$) [43]. For an arbitrary $N$-qubit state $\rho_{s_1 s_2 \ldots s_N}$, a multiqubit entanglement measure can be formulated as [44]

$$\overline{\varrho} = \frac{N}{2} \sum_{1}^{\frac{N}{2}} \varrho_{k|N-k}(\rho_{s_1 s_2 \ldots s_N}) \tag{3-34}$$

where $N$ is assumed even, otherwise $\frac{N}{2}$ should be replaced by $\frac{N-1}{2}$, and $\varrho_{k|N-k}(\rho_{s_1 s_2 \ldots s_N})$ is the entanglement in terms of negativity between two blocks of a bipartition $k|N-k$ of the state $\rho_{s_1 s_2 \ldots s_N}$. We can define the following partition-dependent residual entanglements (PREs)

$$\Pi_{q_1\ldots q_m q_{m+1}\ldots q_k | q_{k+1}\ldots q_n q_{n+1}\ldots q_N} = \varrho^2_{q_1\ldots q_m q_{m+1}\ldots q_k | q_{k+1}\ldots q_n q_{n+1}\ldots q_N}$$

$$-\varrho^2_{q_1\ldots q_m | q_{k+1}\ldots q_n} - \varrho^2_{q_1\ldots q_m | q_{n+1}\ldots q_N} - \varrho^2_{q_{m+1}\ldots q_k | q_{k+1}\ldots q_n} - \varrho^2_{q_{m+1}\ldots q_k | q_{n+1}\ldots q_N} \tag{3-35}$$

and

$$\Pi'_{q_1...q_k|q_{k+1}...q_N} = \varrho^2_{q_1...q_k|q_{k+1}...q_N} - \sum_{i-1}^{k}\sum_{j=k+1}^{N} \varrho^2_{q_i q_j} \qquad (3\text{-}36)$$

For the state in Eq.(3.32), we have

$$\Pi_{q_1...q_m q_{m+1}...q_k|q_{k+1}...q_n q_{n+1}...q_N} = \Pi'_{q_1...q_k|q_{k+1}...q_N} = \varrho^2_{q_1...q_k|q_{k+1}...q_N} = 4|f|^2|f'|^2 \qquad (3\text{-}37)$$

Another useful entanglement measure was introduced in Refs.[45,46] for $n$-qubit state $|\psi\rangle = \sum_{i=0}^{2^n-1} a_i|i\rangle$ with even $n$, as

$$\tau(\psi) = 2|\chi^*(a,n)| \qquad (3\text{-}38)$$

where

$$\chi^*(a,n) = \sum_{i=0}^{2^{n-2}-1} sgn^*(n,i)(a_{2i}a_{(2^{n-1}-1)-2i} - a_{2i+1}a_{(2^{n-2}-2)-2i}), \qquad (3\text{-}39)$$

$$sgn^*(n,i) = \begin{cases} (-1)^{N(i)} & 0 \leq i \leq 2^{n-3}-1 \\ (-1)^{N(i)+n} & 2^{n-3} \leq i \leq 2^{n-2}-1 \end{cases} \qquad (3\text{-}40)$$

where, $N(i)$ is the number of the occurrences of 1 in the $n$-bit binary representation of $i$ as $i_{n-1}...i_1 i_0$ ( in binary representation, $i$ is written as $i = i_{n-1}2^{n-1} + ... + i_1 2^1 + i_0 2^0$). For the state Eq.(3.32), one can see that

$$\tau(\psi) = 2|\chi^*(a,n)| = 2|a_{2^m-1}a_{2^{2m}-2^m}| = 2|ff'|. \qquad (3\text{-}41)$$

In order to achieve maximal entanglement ($GHZ$ state), we should have

$$|f| = |f'| = \frac{1}{\sqrt{2}} \qquad (3\text{-}42)$$

Then $Q(|\psi\rangle) = \Pi_{q_1...q_m q_{m+1}...q_k|q_{k+1}...q_n q_{n+1}...q_N} = \Pi'_{q_1...q_k|q_{k+1}...q_N} = \tau(\psi) = 1$.

In the following we consider the four qubit state (the case $m=2$) $|\psi(t=0)\rangle = |1100\rangle$ and the six qubit state (the case $m=3$) $|\psi(t=0)\rangle = |111000\rangle$ in details: From Eq. (2-12), for $m=2$, the QD parameters are given by

$$\alpha_1 = 2, \quad \alpha_2 = 0; \quad \omega_1 = \omega_2 = 4,$$

Then by using the recursion relations (B-48) and (B-51), we obtain

$$Q_2^{(1)}(x) = x^2 - 2x - 4, \quad Q_3(x) = x(x-4)(x+2),$$

so that the stieltjes function is given by

$$G_\mu(x) = \frac{Q_2^{(1)}(x)}{Q_3(x)} = \frac{x^2 - 2x - 4}{x(x-4)(x+2)}.$$

Then the corresponding spectral distribution is given by

$$\mu(x) = \sum_{l=0}^{2} \gamma_l \delta(x - x_l) = \frac{1}{6}\{3\delta(x) + \delta(x-4) + 2\delta(x+2)\},$$

which indicates that

$$W = \begin{pmatrix} \gamma_0 & 0 & 0 \\ 0 & \gamma_1 & 0 \\ 0 & 0 & \gamma_2 \end{pmatrix} = \frac{1}{6}\begin{pmatrix} 1 & 0 & 0 \\ 0 & 3 & 0 \\ 0 & 0 & 2 \end{pmatrix}.$$

In order to obtain the suitable coupling constants, we need also the eigenvalue matrix $P$ with entries $P_{ij} = P_i(x_j) = \frac{1}{\sqrt{\omega_1 \ldots \omega_i}} Q_i(x_j)$. By using the recursion relations (B-48), one can obtain $P_0(x) = 1$, $P_1(x) = \frac{x}{2}$ and $P_2(x) = \frac{1}{4}(x^2 - 2x - 4)$, so that

$$P = \begin{pmatrix} 1 & 1 & 1 \\ 2 & 0 & -1 \\ 1 & -1 & 1 \end{pmatrix}.$$

Then, Eq. (3-30) leads to

$$-t(J_0 + 2J_1 + J_2) = \theta \pm \frac{\pi}{4} \pm 2c_0\pi,$$

$$-t(J_0 - J_2) = \theta \mp \frac{\pi}{4} \pm 2c_1\pi,$$

$$-t(J_0 - J_1 + J_2) = \theta \pm \frac{\pi}{4} \pm 2c_2\pi.$$

Now, by considering $c_0 = c_1 = c_2 = 0$ we obtain

$$J_0 = -\frac{\theta}{t}, \quad J_1 = 0, \quad J_2 = \mp\frac{\pi}{4t}.$$

Also by considering $c_0 = 0, c_1 = c_2 = 1$ the coupling constants will be

$$J_0 = -\frac{3\theta \pm 5\pi}{3t}, \quad J_1 = \pm\frac{2\pi}{3t}, \quad J_2 = \pm\frac{\pi}{12t}$$

and by considering $c_0 = 1, c_1 = c_2 = 0$

$$J_0 = -\frac{3\theta \pm \pi}{3t}, \quad J_1 = \mp\frac{2\pi}{3t}, \quad J_2 = \mp\frac{7\pi}{12t}$$

From Eq. (2-12), for $m = 3$, the QD parameters are given by

$$\alpha_1 = 4, \quad \alpha_2 = 4, \quad \alpha_3 = 0; \quad \omega_1 = \omega_3 = 9, \quad \omega_2 = 16$$

Then by using the recursion relations (B-48) and (B-51), we obtain

$$Q_3^{(1)}(x) = x^3 - 8x^2 - 9x + 36, \quad Q_4(x) = (x^2 - 9)(x - 9)(x + 1),$$

so that the stieltjes function is given by

$$G_\mu(x) = \frac{Q_3^{(1)}(x)}{Q_4(x)} = \frac{x^3 - 8x^2 - 9x + 36}{(x^2 - 9)(x - 9)(x + 1)}.$$

Then the corresponding spectral distribution is given by

$$\mu(x) = \sum_{l=0}^{3} \gamma_l \delta(x - x_l) = \frac{1}{20}\{\delta(x - 9) + 5\delta(x - 3) + 9\delta(x + 1) + 5\delta(x + 3)\},$$

which indicates that

$$W = \begin{pmatrix} \gamma_0 & 0 & 0 & 0 \\ 0 & \gamma_1 & 0 & 0 \\ 0 & 0 & \gamma_2 & 0 \\ 0 & 0 & 0 & \gamma_3 \end{pmatrix} = \frac{1}{20}\begin{pmatrix} 1 & 0 & 0 & 0 \\ 0 & 5 & 0 & 0 \\ 0 & 0 & 9 & 0 \\ 0 & 0 & 0 & 5 \end{pmatrix}.$$

By using the recursion relations (B-48), one can obtain $P_0(x) = 1$, $P_1(x) = \frac{x}{3}$, $P_2(x) = \frac{1}{12}(x^2 - 4x - 9)$ and $P_3(x) = \frac{1}{36}(x^3 - 8x^2 - 9x + 36)$, so that

$$P = \begin{pmatrix} 1 & 1 & 1 & 1 \\ 3 & 1 & -\frac{1}{3} & -1 \\ 3 & -1 & -\frac{1}{3} & 1 \\ 1 & -1 & 1 & -1 \end{pmatrix}.$$

Then, Eq. (3-30) gives

$$-t(J_0 + 3J_1 + 3J_2 + J_3) = \theta \pm \frac{\pi}{4} \pm 2c_0\pi,$$

$$-t(J_0 + J_1 - J_2 - J_3) = \theta \mp \frac{\pi}{4} \mp 2c_1\pi,$$

$$-t(J_0 - \frac{1}{3}J_1 - \frac{1}{3}J_2 + J_3) = \theta \pm \frac{\pi}{4} \pm 2c_2\pi,$$

$$-t(J_0 - J_1 + J_2 - J_3) = \theta \mp \frac{\pi}{4} \pm 2c_3\pi.$$

Again, by considering $c_0 = c_1 = c_2 = c_3 = 0$ we obtain

$$J_0 = -\frac{\theta}{t}, \quad J_1 = J_2 = 0, \quad J_3 = \frac{\mp\pi}{4t}.$$

## 4  Conclusion

A $2m$-qubit initial state was prepared to evolve under a particular spin Hamiltonian, which could be written in terms of the adjacency matrix of the Johnson graph $J(2m, m)$. By using spectral analysis methods and employing algebraic structures of the Johnson networks, such as distance-regularity and stratification, a method for finding a suitable set of coupling constants in the Hamiltonians associated with the networks was given so that in the final state, the maximal entanglement of the form $GHZ$ state, could be generated. In this work we imposed a constraint so that all amplitudes in the final state were equal to zero except to two amplitudes corresponding to the first and the final strata (any pair of antinodes of the network), where for $J(2m, m)$ these strata contain only one vertex, then $GHZ$ state was generated. We hope to generalize this method to arbitrary Johnson networks $J(n, m)$ and other various graphs, in order to investigate the entanglement of such systems by using some multipartite entanglement measures.

# Appendix

## A  Stratification technique

In this section, we recall the notion of stratification for a given graph $\Gamma$. To this end, let $\partial(x, y)$ be the length of the shortest walk connecting $x$ and $y$ for $x \neq y$. By definition $\partial(x, x) = 0$ for all $x \in V$. The graph becomes a metric space with the distance function $\partial$. Note that $\partial(x, y) = 1$ if and only if $x \sim y$. We fix a vertex $o \in V$ as an origin of the graph, called the reference vertex. Then, the graph $\Gamma$ is stratified into a disjoint union of strata (with respect to the reference vertex $o$) as

$$V = \bigcup_{i=0}^{\infty} \Gamma_i(o), \quad \Gamma_i(o) := \{\alpha \in V : \partial(\alpha, o) = i\} \tag{A-43}$$

Note that $\Gamma_i(o) = \emptyset$ may occur for some $i \geq 1$. In that case we have $\Gamma_i(o) = \Gamma_{i+1}(o) = ... = \emptyset$. With each stratum $\Gamma_i(o)$ we associate a unit vector in $l^2(V)$ defined by

$$|\phi_i\rangle = \frac{1}{\sqrt{\kappa_i}} \sum_{\alpha \in \Gamma_i(o)} |\alpha\rangle, \tag{A-44}$$

where, $\kappa_i = |\Gamma_i(o)|$ is called the $i$-th valency of the graph ($\kappa_i := |\{\gamma : \partial(o, \gamma) = i\}| = |\Gamma_i(o)|$).

One should notice that, for distance regular graphs, the above stratification is independent of the choice of reference vertex and the vectors $|\phi_i\rangle, i = 0, 1, ..., d-1$ form an orthonormal basis for the so called Krylov subspace $K_d(|\phi_0\rangle, A)$ defined as

$$K_d(|\phi_0\rangle, A) = \text{span}\{|\phi_0\rangle, A|\phi_0\rangle, \cdots, A^{d-1}|\phi_0\rangle\}. \tag{A-45}$$

Then it can be shown that [25], the orthonormal basis $|\phi_i\rangle$ are written as

$$|\phi_i\rangle = P_i(A)|\phi_0\rangle, \tag{A-46}$$

where $P_i = a_0 + a_1 A + ... + a_i A^i$ is a polynomial of degree $i$ in indeterminate $A$ (for more details see for example [25,26]).

## B  Spectral distribution associated with the graphs

In this section we recall some facts about spectral techniques used in the paper. From orthonormality of the unit vectors $|\phi_i\rangle$ given in Eq.(A-44) (with $|\phi_0\rangle$ as unit vector assigned to the reference node) we have

$$\delta_{ij} = \langle \phi_i | \phi_j \rangle = \int_R P_i(x) P_j(x) \mu(dx). \tag{B-47}$$

By rescaling $P_k$ as $Q_k = \sqrt{\omega_1 \ldots \omega_k} P_k$, the spectral distribution $\mu$ under question will be characterized by the property of orthonormal polynomials $\{Q_k\}$ defined recurrently by

$$Q_0(x) = 1, \quad Q_1(x) = x,$$

$$xQ_k(x) = Q_{k+1}(x) + \alpha_k Q_k(x) + \omega_k Q_{k-1}(x), \quad k \geq 1. \tag{B-48}$$

The parameters $\alpha_k$ and $\omega_k$ appearing in (B-48) are defined by

$$\alpha_0 = 0, \quad \alpha_k = \kappa - b_k - c_k, \quad \omega_k \equiv \beta_k^2 = b_{k-1} c_k, \quad k = 1, ..., d, \tag{B-49}$$

where, $\kappa \equiv \kappa_1$ is the degree of the networks and $b_i$'s and $c_i$'s are the corresponding intersection numbers. Following Refs. [34], we will refer to the parameters $\alpha_k$ and $\omega_k$ as $QD$ (Quantum Decomposition) parameters (see Refs. [26,27,28,34] for more details). If such a spectral distribution is unique, the spectral distribution $\mu$ is determined by the identity

$$G_\mu(x) = \int_R \frac{\mu(dy)}{x-y} = \cfrac{1}{x - \alpha_0 - \cfrac{\omega_1}{x - \alpha_1 - \cfrac{\omega_2}{x - \alpha_2 - \cfrac{\omega_3}{x - \alpha_3 - \cdots}}}} = \frac{Q_d^{(1)}(x)}{Q_{d+1}(x)} = \sum_{l=0}^d \frac{\gamma_l}{x - x_l}, \tag{B-50}$$

where, $x_l$ are the roots of the polynomial $Q_{d+1}(x)$. $G_\mu(x)$ is called the Stieltjes/Hilbert transform of spectral distribution $\mu$ and polynomials $\{Q_k^{(1)}\}$ are defined recurrently as

$$Q_0^{(1)}(x) = 1, \quad Q_1^{(1)}(x) = x - \alpha_1,$$

$$xQ_k^{(1)}(x) = Q_{k+1}^{(1)}(x) + \alpha_{k+1} Q_k^{(1)}(x) + \omega_{k+1} Q_{k-1}^{(1)}(x), \quad k \geq 1, \tag{B-51}$$

respectively. The coefficients $\gamma_l$ appearing in (B-50) are calculated as

$$\gamma_l := \lim_{x \to x_l}[(x - x_l)G_\mu(x)] \tag{B-52}$$

Now let $G_\mu(z)$ is known, then the spectral distribution $\mu$ can be determined in terms of $x_l, l = 1, 2, ...$ and Gauss quadrature constants $\gamma_l, l = 1, 2, ...$ as

$$\mu = \sum_{l=0}^{d} \gamma_l \delta(x - x_l) \tag{B-53}$$

(for more details see Refs. [35,36,37,38]).